\documentclass[epj]{svjour}

\usepackage{graphicx}
\usepackage{fancyhdr}
\def\makeheadbox{{
 }}
\def\mytitle{My title} 
\def\myauthors{My name}  
\def\mytype{My type of session}
\def\mysession{My session}

\def\mytitle{Studies of CP-conserving and CP-violating $B_{s}$ mixing
parameters with the D0 experiment} 
\def\myauthors{James Walder}    
\def\mytype{Contributed Talk}    
\def\mysession{Flavor Physics}

\newcommand{\bs}        {\ensuremath{B_s^0}}
\newcommand{\ds}        {\ensuremath{D_s^{(*)}}}
\newcommand{\brbs}             {\ensuremath{\mbox{Br}(\bs \to \ds \ds)} }

\newcommand{\ket}[1]               {\ensuremath{\left |{#1}\right>} }

\begin{document}
\title{Studies of CP-conserving and CP-violating $B_{s}$ mixing
parameters with the D0 experiment}
\author{James Walder (for the D0 Collaboration)
}                     
%
%
\institute{Department of Physics,\\
Lancaster University,\\
Lancaster,\\
LA1 4YB,
UK.
}
%
\date{}
\abstract{
This paper summarises the recent results of the Run IIa D0 experiment at the
Tevatron Collider at Fermilab on the observable parameters of the $B_{s}$ meson. 
A measurement of the  branching fraction $\brbs$ is reported, which  provides an
estimate of the width difference $\Delta\Gamma_{s}^{CP}/\Delta\Gamma_{s}$.
Through the decay $B_{s} \to J/\psi \phi$ the width difference
$\Delta\Gamma_{s}$ is extracted,
and for the first time a constraint is set on the CP-violating phase $\phi_{s}$,
although a four-fold ambiguity remains. This result is combined with other D0
measurements to yield $\Delta\Gamma_{s}=0.13\pm0.09\,{\rm ps^{-1}}$,
$\phi_{s} = -0.70^{+0.47}_{-0.39}$.
} 
\maketitle
\section{Introduction\label{sec:intro}}
The standard model (SM) of particle physics, with three families of quarks
contains the CKM matrix which contains one complex phase that governs
CP-violation. 
The $\bs$ system can be described by the  Schr\"{o}dinger equation:
\begin{eqnarray}
i \frac{d}{dt} \!
\left( \!\!\begin{array}{c} B^0_s \\ \bar{B}^0_s
       \end{array}  \!\!\right)\!\!
       =\!\!
\left( \!\!\begin{array}{cc}
       M- \frac{i \Gamma}{2} & M_{12} - \frac{i \Gamma_{12}}{2} \\
       M_{12}^* - \frac{i \Gamma_{12}^*}{2} & M - \frac{i \Gamma}{2}
       \end{array} \!\!\right)\!
\left( \!\!\begin{array}{c} B^0_s \\ \bar{B}^0_s
              \end{array} \!\! \right),\nonumber
\label{eqn:mxingmat}
\end{eqnarray}
where from CPT invariance: $M_{11}=M_{22}\equiv M$, and
$\Gamma_{11}=\Gamma_{22}\equiv\Gamma$. 
The light and heavy mass eigenstates are defined as $\ket{B_{L}}$ and
$\ket{B_{H}}$ respectively, and relate to the flavour eigenstates through
$\ket{B_{L}} = p\ket{B_{s}^{0}} + q\ket{\bar B_{s}^{0}}$, 
$\ket{B_{H}} = p\ket{B_{s}^{0}} - q\ket{\bar B_{s}^{0}}$, where
$|p|^{2}+|q|^{2} = 1$.
 
Under the assumption of no CP-violation, the light and heavy mass eigenstates
correspond to the CP-even and CP-odd eigenstates respectively.

The mass difference, $\Delta M_{s} = M_{H} - M_{L}\sim 2 |M_{12}|$
is sensitive to the effects of new physics, whereas the CP width difference
$\Delta\Gamma^{CP}_{s} = \Gamma_{\rm even} - \Gamma_{\rm odd}\sim 2|\Gamma_{12}|$
does not provide sensitivity to new physics, as $|\Gamma_{12}|$ is dominated by
tree-level processes. However, the width difference $\Delta \Gamma_{s} =
\Gamma_{L}-\Gamma_{H}=\Delta\Gamma^{CP}_{s}\cos\phi_{s}$ is highly sensitive
to possible effects of new physics through the CP-violating phase angle
$\phi_{s}=\arg\left(-\frac{M_{12}}{\Gamma_{12}}\right)$, which in the SM is
expected to be small ($\approx 0$), but may be enhanced through fourth
generation models to $-\phi_{s}\sim 0.5$--$0.7$~\cite{Hou:2006mx}.

The D0 detector is a general purpose spectrometer and
calorimeter~\cite{dzerodet}.
The significant components for these measurements are the muon chambers,
calorimeters and central tracking region. Enclosed within a 2 T superconducting
solenoid is a silicon micostrip tracker (SMT) and central fiber tracker (CFT)
for vertexing and
tracking of charged particles that extends out to a pseudorapidity of $|\eta|=2.0$,
where $\eta=-\ln[\tan(\theta/2)]$, and $\theta$ is the polar angle.
The three liquid-argon/uranium calorimeters provide coverage up to
$|\eta|\approx4.0$.  The muon system consists of one tracking layer and
scintillation trigger counters in front of 1.8~T iron toroids with two layers
after the toroids.  Coverage extends to  $|\eta|=2.0$.

Results given here correspond to data samples recorded by the D0 detector of
integrated luminosities $1.0$--$1.3\,$fb$^{-1}$.

\section{Mass difference $\Delta M_{s}$\label{sec:deltam}}
In 2006 the D0 experiment made the first direct double-sided
constraint~\cite{d0mixing} on the
oscillation frequency $\Delta M_{s}$ of the $B_{s}$ meson in the semileptonic
decays\footnote{Charge-conjugate states are implied throughout.} of $\bs\to
D_{s}\mu\nu X$, $D_{s}\to \phi \pi$, $\phi\to K^{+}K^{-}$. 
A limit of $17<\Delta M_{s}<21\,$ps$^{-1}$ at $90\%$ CL was measured. A 
more recent measurement~\cite{cdfmixing}, which includes hadronic modes of
$B_{s}$ meson decay, gives the precision value $\Delta M_{s}=17.77\pm
0.10\,({\rm stat})\pm 0.07 ({\rm syst})\,$ps$^{-1}$.

\section{Width difference $\Delta\Gamma_{s}$\label{sec:deltagamma}}

\subsection{$\brbs$}
The decay of $\bs$ mesons to $D_{s}^{+}D_{s}^{-}$ produces a CP-even final
state~\cite{cpeven}, and under certain theoretical assumptions, the decay $\bs\to\ds\ds$
is also predominately CP-even, up to\footnote{It
should be noted that some estimates consider a CP-odd fraction up to $30\%$.}
$\sim95\%$~\cite{dsdseven}.
Under these assumptions, the branching fraction of this decay can be related to
the CP width difference 
$\Delta\Gamma_{s}^{CP}\approx 2|\Gamma_{12}|$ 
$(\Delta\Gamma_{s} = \Delta\Gamma_{s}^{CP}\cos\phi_{s})$ by:
\begin{eqnarray}
2 \brbs 
=
\frac{\Delta\Gamma_{s}^{CP}}{\Gamma_{s}}
  \left[
    1 + {\mathcal O}\left(\frac{\Delta\Gamma_{s}^{CP}}{\Gamma_{s}}\right)
   \right].
\label{eqn:dgdsds}
\end{eqnarray}

The D0 experiment performed a measurement~\cite{bsdsds} using the decay chain $\bs\to
D_{s}^{(*)\pm} D_{s}^{(*)\mp}$, $D_{s}^{\pm}\to\phi\pi^{\pm}$,
$D_{s}^{\mp}\to\phi\mu^{\mp}\nu X$, $\phi\to K^{+}K^{-}$ to extract the
branching fraction $\brbs$ and the width difference
$\Delta\Gamma^{CP}_{s}/\Gamma_{s}$.

To reduce detector related systematics effects, the above process was
normalised to to the decay $\bs\to D_{s}^{(*)\pm}\mu^{\mp}\nu X$, 
$D_{s}^{\pm}\to\phi\pi^{\pm}$, $\phi\to K^{+}K^{-}$, which has a similar
final-state to the main process. From a fit to the invariant mass of the $D_{s}(\phi\pi)$
mass peak, $17670\pm 230$ events were estimated for the normalisation data.

\begin{figure}
\includegraphics[width=0.4\textwidth,height=0.4\textwidth,angle=0]{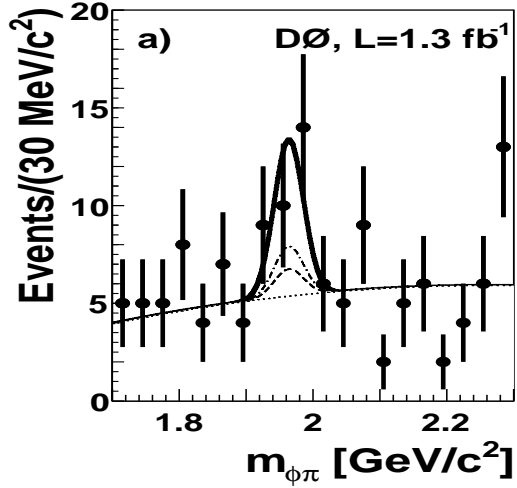}
\caption{Invariant mass distribution of the $D_{s}(\phi\pi)$ meson projected
onto the signal region of the fitting procedure.  The curves are the projected
fit results for the: total fit (solid);  polynomial background contribution
(dotted); non-peaking background component (dashed); background peaking in the
mass region of both the $\phi$ and $D_{s}$ mesons (dash-dotted).}
\label{fig:mds}       
\end{figure}

\begin{figure}
\includegraphics[width=0.4\textwidth,height=0.4\textwidth,angle=0]{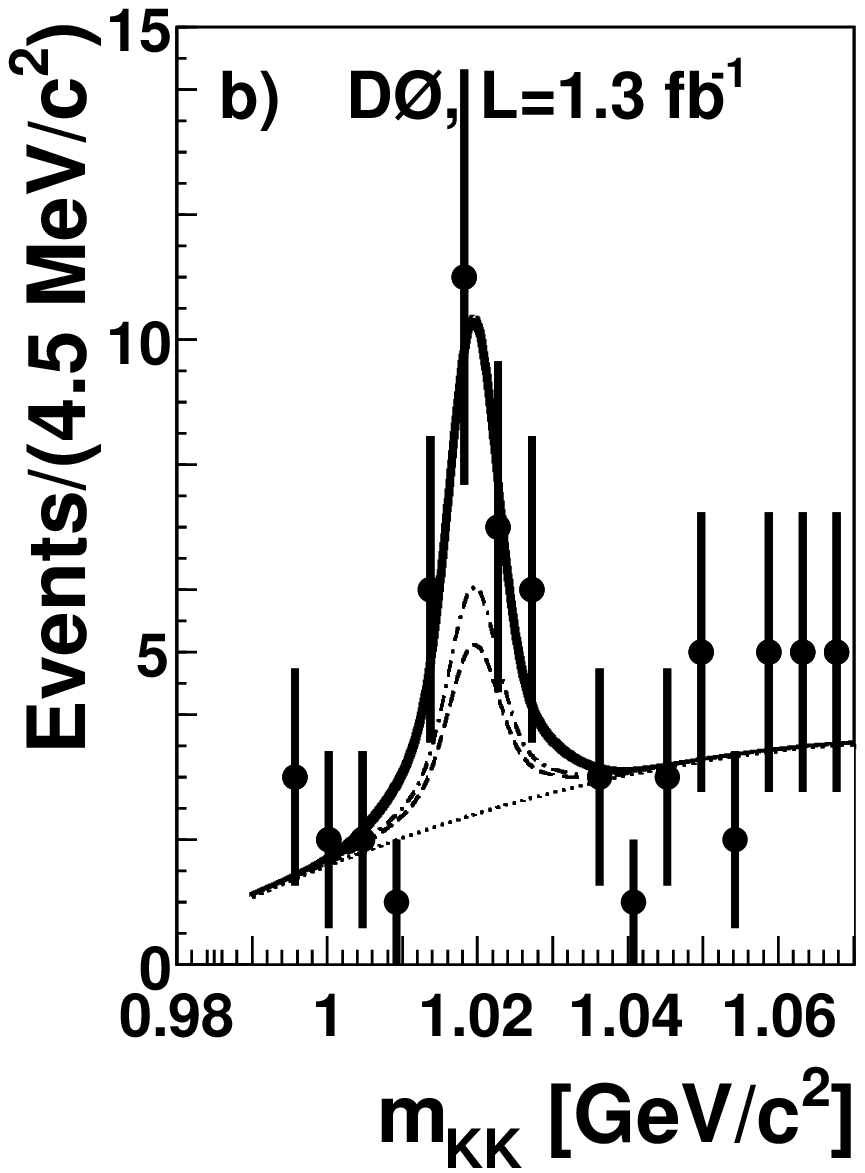}
\caption{Invariant mass distribution of the $\phi$ meson from $D_{s}\to \phi
\mu \nu X$, projected
onto the signal region of the fitting procedure.  The curves are the projected
fit results for the: total fit (solid);  polynomial background contribution
(dotted); non-peaking background component (dashed); background peaking in the
mass region of both the $\phi$ and $D_{s}$ mesons (dash-dotted).}
\label{fig:mkk}       
\end{figure}

For the signal decay, $13.4^{+6.6}_{-6.0}$ events were estimated in data, from a
two-dimensional fit to the invariant mass of $D_{s}$ mesons from
$D_{s}\to\phi\pi$, and $\phi$ mesons in the decay $D_{s}\to\phi\mu\nu X$. 
Figures~\ref{fig:mds} and~\ref{fig:mkk} shows the result of the fitting
procedure projected onto the signal regions of the mass variable not plotted.

Using data and Monte Carlo
(MC) simulations, $2.0$ background events were estimated to contribute to the
signal number. 
The difference between the efficiencies to reconstruct the signal and
normalisation processes is determined in MC, and corrections applied to the
in determination of the result.

The branching fraction $\brbs$ was measured to be:
\begin{eqnarray}
\brbs=0.039^{+0.019}_{-0.017}\,({\rm stat})^{+0.016}_{-0.015}\,({\rm syst}).
\label{eqn:brbsresult}\nonumber
\end{eqnarray}
Under the assumptions leading to Eq.~\ref{eqn:dgdsds}, the width difference is
estimated to be:
\begin{eqnarray}
\frac{\Delta\Gamma_{s}^{CP}}{\Gamma_{s}} = 0.079^{+0.038}_{-0.035}\,({\rm
stat})^{+0.031}_{-0.030}\,({\rm syst}).
\label{eqn:brbsdgresult}
\end{eqnarray}

\subsection{$\bs\to J/\psi \phi$}
The decay $\bs\to J/\psi \phi$ contains final states with both CP-even and
CP-odd components, which may be separated using a lifetime-dependent
angular analysis, leading to a measurement of the lifetime difference. If the
lifetime difference is sufficiently large, the CP-violating phase $\phi_{s}$
can also be extracted.

The measurement~\cite{jpsiphi} was performed in the decay of untagged $B_{s}$
mesons: $\bs\to J/\psi\phi$, $J/\psi \to \mu^{+}\mu^{-}$, $\phi\to K^{+}K^{-}$
with a set of data corresponding to an  integrated luminosity of $1.1\,$fb$^{-1}$.
From $23300$ events
in the final selection $1039\pm45$ were estimated from the fitting procedure to
originate from the $B_{s}$ decay. The likelihood fitting procedure uses the
angle between the kaon and $J/\psi$ in the $\phi$ meson rest frame, and the
tranversity polar and azimuthal angles of the muon in the $J/\psi$ rest frame
to separate out the CP-eigenstates.
The background is separated into contributions of: prompt, from directly
produced $J/\psi$ mesons and additional track from hadronisation; and
non-prompt, where $J/\psi$ mesons  are produced from $B$ mesons decays, but
combined with a $\phi$ meson from track from hadronisation or multi-body
decays of the $B$ meson.

\begin{figure}
\includegraphics[width=0.45\textwidth,height=0.4\textwidth,angle=0]{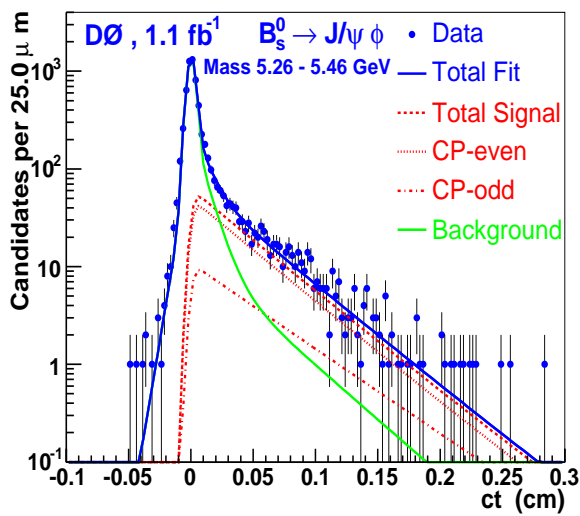}
\caption{Proper decay-length of the $B_{s}$ meson candidates in $B_{s}\to J\psi
\phi$. The curves describe: the total fit in blue (solid); the total signal
contribution in red (dashed) with the CP-even (dotted) and CP-odd
(dash-dotted) separated; and the background in green (lower solid).}
\label{fig:jpsilifetime}       
\end{figure}

In Fig.~\ref{fig:jpsilifetime} the contributions of the two CP-eigenstates and
backgrounds are shown for the proper decay-lengths of $B_{s}$ meson candidates. 

Under the assumption of no CP-violation $(\phi_{s}\equiv 0)$, the maximum
likelihood fit yields
\begin{eqnarray}
|\Delta\Gamma_{s}| &=& 0.12^{+0.08}_{-0.10}\,{\rm ps^{-1}},\nonumber\\
\tau_{s}       &=&  1.52\pm0.08\,{\rm ps}.\nonumber
\label{eqn:jpsiconserveddg}
\end{eqnarray}

\section{CP-violating phase $\phi_{s}$\label{sec:phi}}

If there exists a sizeable width difference in the $\bs$ meson system, there is
sensitivity to the CP-violating phase angle through the untagged
time-dependent width 
$\Gamma_{s}(t)\sim(e^{-\Gamma_{s}t}-e^{\Gamma_{H}t})\sin\phi_{s}$.

From the same $\bs\to J/\psi \phi$ analysis described above, the constraint on
$\phi_{s}$ was removed and this parameter allowed to float, and the fitting
procedure was repeated.
In this case there exists a four-fold ambiguity in the results, due to a  
sign reversal of $\sin\phi_{s}$ with the simultaneous
flip of the strong phase angles $\cos\delta_{1}$ and $\cos\delta_{2}$ which
appears in the likelihood fitting procedure.
The average lifetime extracted from the fit is
\begin{eqnarray}
\tau_{s} &=& 1.49\pm0.08\,{\rm ps},\nonumber
\end{eqnarray}
and for the case $\cos\delta_{1}< 0$, $\cos\delta_{2}> 0$: 
\begin{eqnarray}
|\Delta\Gamma_{s}|&=& 0.17^{+0.09}_{-0.09}\,{\rm ps^{-1}},\nonumber\\
\phi_{s}&=& \left\{
    \begin{array}{lc}-0.79 \pm 0.56,&\;\Delta\Gamma_{s}>0,\\
                            +2.35 \pm 0.56,&\;\Delta\Gamma_{s}<0,\\
    \end{array}\right.\nonumber
\end{eqnarray}
and in the case  $\cos\delta_{1}> 0$, $\cos\delta_{2}< 0$:
\begin{eqnarray}
|\Delta\Gamma_{s}|&=& 0.17^{+0.09}_{-0.09}\,{\rm ps^{-1}},\nonumber\\
\phi_{s}&=& \left\{
    \begin{array}{lc}+0.79 \pm 0.56,&\;\Delta\Gamma_{s}>0,\\
                            -2.35 \pm 0.56,&\;\Delta\Gamma_{s}<0.\\
    \end{array}\right.\nonumber
\end{eqnarray}

\section{Combination results\label{sec:combination}}

Using the results from the $\bs\to J/\psi \phi$ analysis with the $\Delta
M_{s}$ mixing measurement~\cite{cdfmixing} and the world-average flavour 
specific lifetime, which includes a D0 measurment~\cite{d0wafstau}, and an
additional constraint described below,
an improved estimate of the CP-violating phase value was obtained~\cite{combination}.

Two charge asymmetry measurements by the D0 experiment are used in forming the
additional constraint. 
In Ref.~\cite{bsmumu}, the same-sign di-muon charge asymmetry, defined as 
\[
A_{SL}^{\mu\mu}=
\frac{N(b\bar{b}\to\mu^{+}\mu^{+}X) - N(b\bar{b}\to\mu^{-}\mu^{-}X)}
         {N(b\bar{b}\to\mu^{+}\mu^{+}X) + N(b\bar{b}\to\mu^{-}\mu^{-}X)}
\]
was measured, and from this asymmetry value $A_{SL}^{s}$ extracted:
\begin{eqnarray}
A_{SL}^{s}= -0.0064\pm0.0101.
\label{eqn:aslmumu}
\end{eqnarray}

In Ref.~\cite{bsmuds} the charge asymmetry value from semileptonic decays from
$\bs\to D_{s} \mu \nu$, $D_{s}\to \phi\pi$ was measured from which the value
$A_{SL}^{s}$ extracted:
\begin{eqnarray}
A_{SL}^{s} = +0.0245\pm0.0193\,{\rm (stat)}\pm0.0035\,{\rm (syst)}.
\label{eqn:asldsmu}
\end{eqnarray}

Together, Eqs.~\ref{eqn:asldsmu} and~\ref{eqn:aslmumu} are almost independent
and can be combined to provide the current best limits on $A_{SL}^{s}$:
\begin{eqnarray}
A_{SL}^{s}= 0.0001\pm0.0090.
\label{eqn:aslcombined}
\end{eqnarray}
This charge asymmetry value, with the measurement~\cite{cdfmixing} of $\Delta
M_{s}$, is used to provide the additional constraint~\cite{Beneke:2003az}:
\begin{eqnarray}
\Delta\Gamma_{s}\cdot \tan\phi_{s} = A_{SL}^{s}\cdot \Delta M_{s}
=0.02\,{\rm ps^{-1}}.
\label{eqn:tanasl}
\end{eqnarray}

\begin{figure}
\includegraphics[width=0.45\textwidth,height=0.45\textwidth,angle=0]
{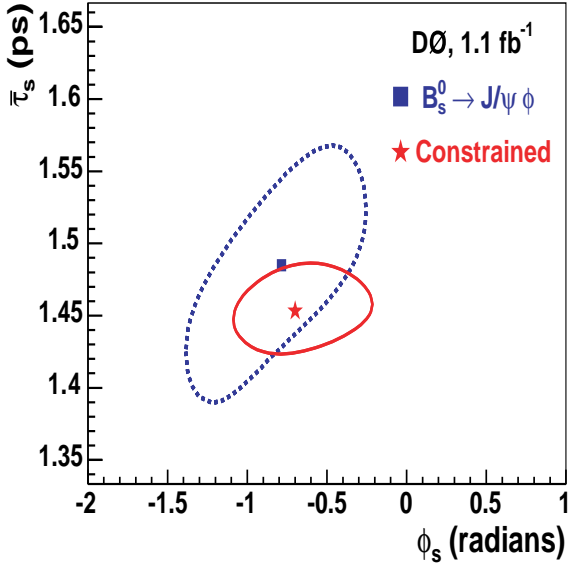}
\caption{Contours of $\ln{\mathcal L}=0.5$ ($39\%$ CL) in the plane of
lifetime vs. CP-violating phase angle. In blue (dotted) the result from the
$\bs\to J/\psi \phi$ analysis, and in red (solid) the combination analysis
results, both shown for the sign combination whose result is closest to the SM
prediction.}
\label{fig:lifetimevsphi}       
\end{figure}

\begin{figure}
\includegraphics[width=0.45\textwidth,height=0.45\textwidth,angle=0]
{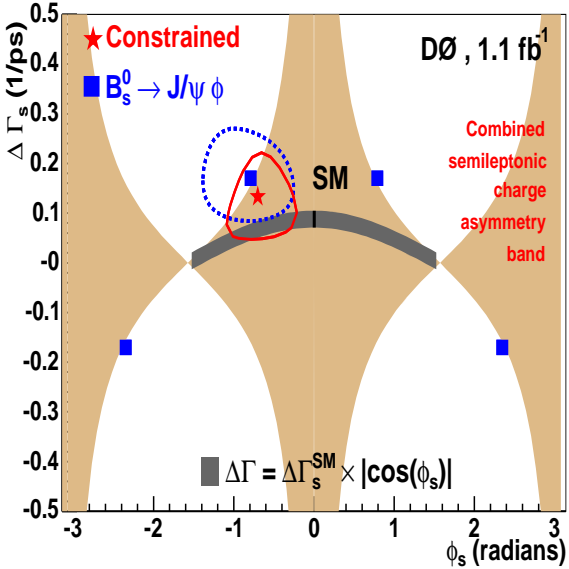}
\caption{Contours of $\ln{\mathcal L}=0.5$ ($39\%$ CL) in the plane of
width difference vs. CP-violating phase angle for solutions with sign
combination closest to the SM prediction of the $\bs\to J/\psi \phi$ analysis
(blue, dotted), and the results of the combination analysis (red, solid).
The four solid (blue) squares represents the central values from the $\bs\to J/\psi
\phi$ showing the four-fold ambiguity.
The SM prediction is shown as the black vertical bar, and the dark band is the
result of $\Delta\Gamma_{s} = \Delta\Gamma_{s}^{SM}|\cos\phi_{s}|$,
$\Delta\Gamma_{s}^{SM} = 0.088 \pm 0.017\,$ps$^{-1}$~\cite{ul}.
The lighter shaded area corresponds to the result of Eq.~\ref{eqn:tanasl}.
}
\label{fig:dgammavsphi}       
\end{figure}

The fit to the $\bs\to J/\psi \phi$ data was repeated with this constraint, and
the results of which are shown as projections in Fig.~\ref{fig:lifetimevsphi}
for the contour of $\tau_{s}$ versus $\phi_{s}$ and in
Fig.~\ref{fig:dgammavsphi} for $\Delta\Gamma_{s}$ versus $\phi_{s}$.
The contours are for the change in likelihood value $\Delta \ln ({\mathcal
L})=0.5$, which corresponds to a $39\%$ CL.
Whilst the four-fold ambiguity on the sign of $\Delta\Gamma_{s}$ with
$\phi_{s}$ remains, the solution closest to the SM prediction yields:
\begin{eqnarray}
\Delta\Gamma_{s}=0.13\pm0.09\,{\rm ps^{-1}},\nonumber\\
\phi_{s} = -0.70^{+0.47}_{-0.39}.\nonumber
\end{eqnarray}

\section{Summary\label{sec:summary}}

The $B_{s}$ meson now has information regarding all its observable parameters,
be that in precision measurements (i.e. $M_{s}$, $\Delta M_{s}$), or a
constraint, such as in $\phi_{s}$.

Whilst no evidence for beyond standard model effects are present, with
increased luminosity and improvements in analysis techniques, the
available parameter space in which it may manifest is certainly diminishing.


%

\begin{thebibliography}{999}
%
%
\bibitem{Hou:2006mx}
  W.~S.~Hou, M.~Nagashima and A.~Soddu,
  Phys.\ Rev.\  D {\bf 76} (2007) 016004, arXiv:hep-ph/0610385.
\bibitem{dzerodet} V.~M.~Abazov {\it et al.}  [D0 Collaboration],
 Nucl. Instrum. Meth. A {\bf 565} (2006) 463.
\bibitem{d0mixing}
  V.~M.~Abazov {\it et al.}  [D0 Collaboration],
  Phys.\ Rev.\ Lett.\  {\bf 97} (2006) 021802.
\bibitem{cdfmixing}
  A.~Abulencia {\it et al.}  [CDF Collaboration],
  Phys.\ Rev.\ Lett.\  {\bf 97} (2006) 242003.
  
  
 \bibitem{cpeven}
I.~Dunuetz {\it et al.}, Phys. Rev. D {\bf 63}, (2001) 114015.
 \bibitem{dsdseven}
  R.~Alexan {\it et al.}, Phys. Lett. B {\bf 316}, (1993) 567.
 \bibitem{bsdsds}
 V.~M.~Abazov {\it et al.}  [D0 Collaboration],
  arXiv:hep-ex/0702049.
 
 \bibitem{jpsiphi}
  V.~M.~Abazov {\it et al.}  [D0 Collaboration],
  Phys.\ Rev.\ Lett.\  {\bf 98}, (2007) 121801.
 
 \bibitem{d0wafstau}
 V.~M.~Abazov {\it et al.} [D0 Collaboration],
 Phys. Rev. Lett. {\bf 97}, (2006)  241801.

 \bibitem{combination}
  V.~M.~Abazov {\it et al.}  [D0 Collaboration],
  arXiv:hep-ex/0702030. 
 \bibitem{bsmumu}
  V.~M.~Abazov {\it et al.}  [D0 Collaboration],
  Phys.\ Rev.\  D {\bf 74}, (2006) 092001.
  \bibitem{bsmuds} 
  V.~M.~Abazov {\it et al.}  [D0 Collaboration],
  Phys.\ Rev.\ Lett.\  {\bf 98}, (2007) 151801.
\bibitem{Beneke:2003az}
  M.~Beneke, G.~Buchalla, A.~Lenz and U.~Nierste,
  Phys.\ Lett.\  B {\bf 576}, (2003) 173.  
  
  \bibitem{ul}
  A.~Lenz and U.~Nierste, hep-ph/0612167.
  
  \end{thebibliography}
%

\end{document}